\documentclass[aps,prl,reprint,superscriptaddress, floatfix]{revtex4-2}
\usepackage{tabularx}
\usepackage{amsmath}
\usepackage{amssymb}
\usepackage{graphicx}% Include figure files
\usepackage{dcolumn}% Align table columns on decimal point
\usepackage{bm}% bold math

\begin{document}

%\preprint{}

\title{Longitudinal spin fluctuations driving field-reinforced superconductivity in UTe$_2$}


\author{Y.\,Tokunaga}
%\thanks{}
\affiliation{%
ASRC,
Japan Atomic Energy Agency
Tokai, Ibaraki 319-1195, Japan
}%
\author{H.\,Sakai} 
\affiliation{%
ASRC,
Japan Atomic Energy Agency
Tokai, Ibaraki 319-1195, Japan
}%
\author{S.\,Kambe} \affiliation{%
ASRC,
Japan Atomic Energy Agency
Tokai, Ibaraki 319-1195, Japan
}%
\author{P.\,Opletal} \affiliation{%
ASRC,
Japan Atomic Energy Agency
Tokai, Ibaraki 319-1195, Japan
}%
\author{Y.\,Tokiwa} \affiliation{%
ASRC,
Japan Atomic Energy Agency
Tokai, Ibaraki 319-1195, Japan
}%
\author{Y.\,Haga} \affiliation{%
ASRC,
Japan Atomic Energy Agency
Tokai, Ibaraki 319-1195, Japan
}%
\author{S.\,Kitagawa} \affiliation{%
Department of Physics, Graduate School of Science, 
Kyoto University, Kyoto 606-8502, Japan
}%
\author{K.\,Ishida} \affiliation{%
Department of Physics, Graduate School of Science, 
Kyoto University, Kyoto 606-8502, Japan
}%
\author{D.\,Aoki}
\affiliation{%
IMR, Tohoku University, Ibaraki 311-1313, Japan
}%
\affiliation{%
Univ. Grenoble Alpes, CEA, Grenoble-INP, IRIG, Pheliqs, 38000 Grenoble France
}%
\author{G.\,Knebel}
\affiliation{%
Univ. Grenoble Alpes, CEA, Grenoble-INP, IRIG, Pheliqs, 38000 Grenoble France
}%
\author{G. \,Lapertot}
\affiliation{%
Univ. Grenoble Alpes, CEA, Grenoble-INP, IRIG, Pheliqs, 38000 Grenoble France
}%
\author{S.\,Kr\"{a}mer}
\affiliation{%
Laboratoire National des Champs Magn\'{e}tiques Intenses, LNCMI-CNRS (UPR3228), EMFL, Universit\'{e} \\ Grenoble Alpes, UPS and INSA Toulouse, Bo\^{\i}te Postale 166, 38042 Grenoble Cedex 9, France
}%
\author{M.\,Horvati{\'c}}
\affiliation{%
Laboratoire National des Champs Magn\'{e}tiques Intenses, LNCMI-CNRS (UPR3228), EMFL, Universit\'{e} \\ Grenoble Alpes, UPS and INSA Toulouse, Bo\^{\i}te Postale 166, 38042 Grenoble Cedex 9, France
}%


\date{\today}

\begin{abstract}
Our measurements of $^{125}$Te NMR relaxations reveal an enhancement of electronic spin fluctuations above  $\mu_0H^*\sim15$~T, leading to their divergence in the vicinity of the metamagnetic transition at $\mu_0H_m\approx35$~T, below which field-reinforced superconductivity appears when a magnetic field ($H$) is applied along the crystallographic $b$ axis. 
The NMR data evidence that these fluctuations are dominantly longitudinal, providing a key to understanding the peculiar superconducting phase diagram in $H\|b$, where such fluctuations enhance the pairing interactions.
\end{abstract}

%\keywords{}

\maketitle

The uranium-based superconductor UTe$_2$ provides an attractive platform for studying the novel physics of spin-triplet and topological superconductivity (SC) in bulk materials \cite{Aoki2022Unconventional-}.
The compound displays bulk superconductivity below $T_c$ = $1.6 - 2.1$~K \cite{Ran2019Nearly-ferromag,Rosa2022Single-thermody,Sakai2022Single-crystal-}. The formation of spin-triplet pairing is supported by a tiny reduction in the NMR Knight shift in the SC state \cite{Nakamine2019Superconducting,Fujibayashi2022Superconducting}, large upper critical field exceeding the ordinary Pauli paramagnetic limit  \cite{Ran2019Nearly-ferromag, Ran2019Extreme-magneti,Knebel2019Field-Reentrant,Aoki2019Unconventional-}, and multiple SC phases similar to superfluid $^{3}$He \cite{Braithwaite2019Multiple-superc,Thomas2020Evidence-for-a-,Aoki2020Multiple-Superc,Rosuel2022_Field-induced,Kinjo2022_changeof,sakai2022field}.
The topological properties of the SC excitations, on the other hand, are suggested from measurements of the STM \cite{Jiao2020Chiral-supercon}, Kerr-effect \cite{Hayes2021Multicomponent-,Wei2022}, and London penetration depth \cite{Bae2021Anomalous-norma,Ishihara2021Chiral-supercon} that have detected anomalous effects potentially originating from a chiral SC state, but are still debated.

The upper critical field of the SC in UTe$_2$ increases significantly when a magnetic field ($H$) is applied exactly along the crystallographic $b$ axis, which is perpendicular to the easy magnetic $a$ axis along which the uranium 5$f$ spin moments favor aligning with an Ising character \cite{Ran2019Extreme-magneti,Knebel2019Field-Reentrant,Aoki2019Unconventional-}.
With increasing the $H$, $T_{\rm c}$ initially decreases as in ordinary superconductors but starts to rise above $\mu_0H^{\ast}\sim 15$~T. 
This increase continues up to 35~T, beyond which $T_c$ suddenly drops to zero at higher magnetic fields.
Such a field-reinforced SC behavior is closely related to the field-induced, first-order metamagnetic transition emerging at $\mu_0H_{\rm m} \approx 35$~T in $H\|b$ \cite{Ran2019Extreme-magneti,Knebel2019Field-Reentrant,Knafo2019Magnetic-Field-,Miyake2019Metamagnetic-Tr,Imajo2019Thermodynamic-I,Knafo2021Comparison-of-t,Thebault_UTe2_2022}.
A phase boundary separating a low-field superconducting (LFSC) state from a high-field superconducting (HFSC) state has been recently discovered around $\mu_0H^{\ast}\sim 15$~T \cite{Rosuel2022_Field-induced,Kinjo2022_changeof,sakai2022field}. The $\mu_0H^{\ast}$ has been found to be insensitive to the sample quality, and thus $T_{c}$ at zero field. A possible change of the pairing mechanism has been discussed between the LFSC and HFSC \cite{Rosuel2022_Field-induced}. Moreover, the signature of another phase boundary has been detected inside the LFSC state around 13~T \cite{sakai2022field}. 

Several theoretical models have been proposed to explain such an anomalous SC phases diagram in $H\|b$ \cite{Ishizuka2019_Insulator,Ishizuka2021_Periodic,Shishidou2021Topological-ban,Machida2020Theory-of-Spin-,Machida2020Notes-on-Multip}. While these models predict different SC order parameters for the LFSC and HFSC states, they commonly require some mechanism to boost $T_{\rm c}$ in higher magnetic fields. In most cases, this is assumed to occur through enhanced spin fluctuations that may arise near $H_{\rm m}$.
This assumption is based on the similarities of the SC phase diagram to that for uranium-based ferromagnetic (FM) superconductors, URhGe \cite{Levy1343} and UCoGe \cite{Aoki2009Extremely-Large}. 
In these materials, the SC state is established within the FM state, and the pairing interaction is thought to be mediated by the exchange of FM spin fluctuations \cite{Saxena_UGe2Nature2000,Aoki2001Coexistence-of-,PhysRevLett.99.067006,fay1980coexistence,valls1984superconductivity}. Under magnetic fields applied along the magnetically hard axis, the compounds exhibit field-induced (URhGe) or field-reinforced (UCoGe) SC, similar to UTe$_2$. NMR experiments have revealed that the excitation spectrum of the spin fluctuations is modified strongly by applied field, depending on its strength and direction. \cite{Hattori2012Superconductivi,Tokunaga2015Reentrant-Super,TokunagaY:PRB93:2016,TokunagaY:JPSCP30:2020,Ishida2021Pairing-interac,KotegawaH:JPSJ84:2015}. This implies  that the strength of the pairing interactions would also depend on magnetic fields, and in fact, 
such a field-dependent pairing mechanism well explain many of the unconventional SC phenomena observed under magnetic fields in the FM superconductors
\cite{Hattori2012Superconductivi,Tokunaga2015Reentrant-Super,TokunagaY:PRB93:2016,TokunagaY:JPSCP30:2020,Ishida2021Pairing-interac,KotegawaH:JPSJ84:2015,Wu2017Pairing-mechani,Tada2016, Hattori_Tsunetsugu_PRB_2013,SuzukiHattori2019,Hattori_Suzuki_2020,Mineev_2017}.

In this Letter, we present the results of high-field $^{125}$Te NMR experiments performed on high-quality single crystals of UTe$_2$.
Our NMR measurements with magnetic fields applied along the $b$ of the crystal axis up to 36 T reveal the development of longitudinal $(\|b)$ spin fluctuations above $\mu_0H^{\ast}\sim 15$~T and their divergence near the field-induced metamagnetic transition at $\mu_0 H_m\approx 35$~T.
These findings offer valuable insights into the unique SC phase diagram for $H\|b$, where the diverging longitudinal spin fluctuations enhance the pairing interactions, resulting in the boost of $T_{\rm c}$ above $H^*$.

%%%%
High-field NMR data were obtained using a 24~MW resistive magnet at the LNCMI-Grenoble. The experiment was performed on a $^{125}$Te enriched (99\%) single crystal (\#1) with $T_{\rm c}=$2.0~K grown by the molten salt flux (MSF) method \cite{Sakai2022Single-crystal-}. $^{125}$Te nuclei have the nuclear gyromagnetic ratio $\gamma_N$=13.454~MHz/T ($I$=1/2) with a natural abundance of 7\%, so that the enrichment largely enhances the NMR signal intensity and the signal to noise ratio. 
NMR experiments at lower fields below 15~T were performed using a SC magnet on a single crystal (\#2)  with $T_{\rm c}=$1.5~K grown by the chemical vapor transport (CVT) method \cite{Ran2019Nearly-ferromag}. The $H$ dependence of the NMR spectrum, the nuclear spin-lattice  relaxation rate ($1/T_1$), and the nuclear spin-spin relaxations rate ($1/T_2$) were measured at a fixed temperature of 2.1~K for the \#1 and 1.6~K for the \#2 crystals; these temperatures are about 0.1~K above the zero-field $T_{\rm c}$ of each crystal.

In the orthorhombic structure of UTe$_2$ (space group No. 71, $Immm$, $D_{2h}^{25}$)\cite{Haneveld1970The-crystal-str,Beck1988Die-Verfeinerun} , Te atoms have two crystallographically inequivalent sites, Te(1) and Te(2), in the unit cell. Thus, the $^{125}$Te NMR spectrum consists of two distinct peaks arising from the two inequivalent sites in $H \| b$. In previous studies, we found no qualitative differences for $1/T_1$ and $1/T_2$ behaviors between the two peaks \cite{Tokunaga2019125Te-NMR-Study}. 
We thus focus on the NMR peak at the higher frequency, corresponding to the Te(2) site in this study.

\begin{figure}[tb]
\begin{center}
\includegraphics[width=6.3 cm,keepaspectratio]{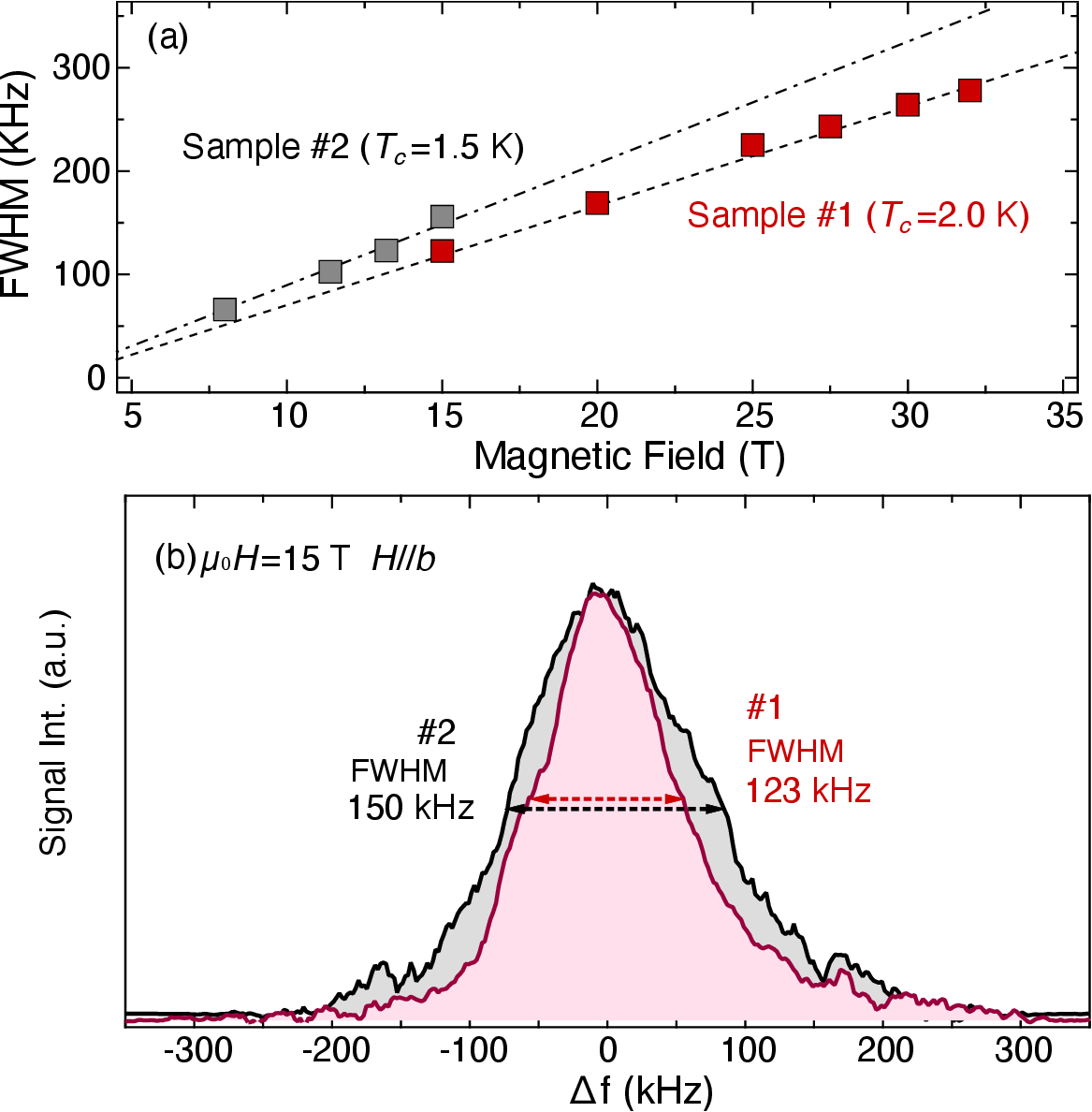} %8.5cm
\caption{(a) Field-dependences of the NMR line width (full width at half maximum) from two different $T_{\rm c}$ samples, \#1 ($T_{\rm c}$=2.0~K) and \#2 ($T_{\rm c}$=1.5~K). (b) Comparison of the NMR spectra obtained in these two crystals in the field of 15~T. }
\end{center}
\label{SPC}
\vspace{-5mm}
\end{figure}

In Fig.\,1 (b), we compared the NMR spectra observed in crystals \#1 and \#2 with different $T_{\rm c}$ in the field of 15~T. 
The NMR linewidth of the \#1  crystal is about 20\% narrower than that of the \#2 crystal. 
The results support microscopically our expectation that crystals with a higher $T_{\rm c}$ have fewer crystal defects and/or less disorder than lower $T_{\rm c}$ crystals \cite{Tokunaga2022Slow-Electronic,Haga2022Effect-of-urani,Sakai2022Single-crystal-}.
As seen in Fig.\,1(a), the NMR linewidth of  both crystals increases linearly with increasing $H$, as expected when the distribution of the NMR shift determines the linewidth.

Figure 2 shows the field dependence of the NMR shift $\Delta f=f_{\rm NMR}-f_{\rm 0}$, where $f_{\rm NMR}$ is the peak frequency of NMR spectrum determined by the peak position of each spectrum, and $f_{\rm 0}=\gamma_N H$. While the NMR line widths are different between the two crystals, there is no significant change in $\Delta f (H)$, so their $\Delta f (H)$ are smoothly connected at 15~T.  As expected, the $\Delta f (H)$ follows $M(H)$, and hence, it is nearly proportional to $H$ up to 32~T. This provides a nearly field-independent Knight shift at $K=\Delta f/f_{\rm 0} \simeq 5.5-6$\%, as shown in the inset of Fig.\,1, whereas the $K(H)$ presents only a very weak, gradual increase. The $K$ values are consistent with those in previous reports \cite{Tokunaga2019125Te-NMR-Study,Kinjo2022_changeof}.
Note that no NMR shift data is above 32~T, although our experiment was performed up to 36~T.  In this field region, we completely lost the NMR spin-echo signals due to extremely short $T_2$, as discussed below.

\begin{figure}[tb]
\begin{center}
\includegraphics[width=6.8 cm,keepaspectratio]{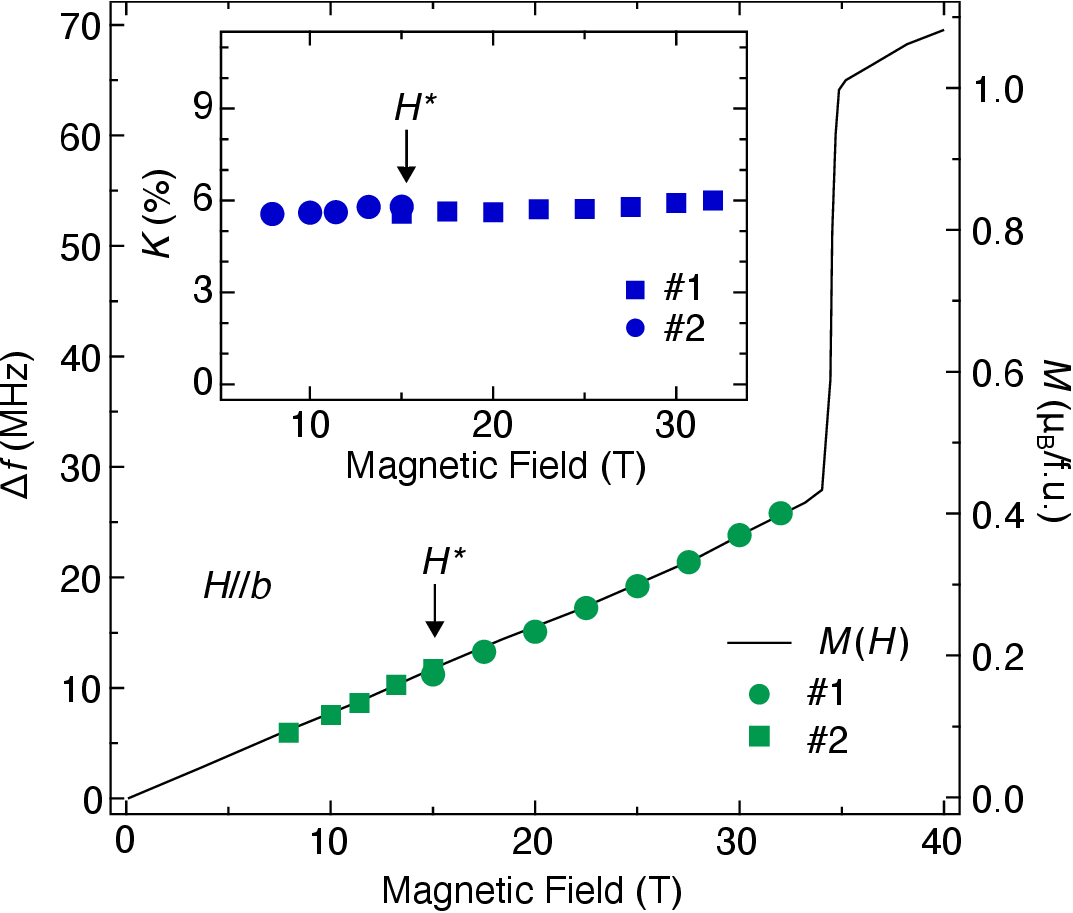} %8.5cm
\caption{The field dependence of the NMR shift $\Delta f=f_{\rm NMR}-f_{\rm 0}$, where $f_{\rm NMR}$ is the peak frequency of NMR spectrum determined by the peak position of each spectrum at 2.1 (1.6)~K for \#1 (\#2) crystals and $f_{\rm 0}=\gamma_N H$. The inset shows the field dependence of the Knight shift, $K=\Delta f/f_{\rm 0}$. The solid line is the magnetization curve at 1.4~K \cite{Miyake2019Metamagnetic-Tr}.}
\end{center}
\label{Knight}
\vspace{-7mm}
\end{figure}

We now turn to the spin dynamics for $H\|b$. 
Figures 3 (a) and (b) show the field dependence of the $1/T_1$ and $1/T_2$ relaxation rates up to 32~T.
Here, $1/T_1$ was measured by applying the saturation $\pi/2$ pulse at time $t$ before the $\pi/2-\tau-\pi$ spin-echo sequence used to record the recovery data $R(t)$, and was evaluated by fitting $R(t)$ to the exponential function for spin $I=1/2$ nuclei ($^{125}$Te)  \cite{NarathA:PR162:1967}. $1/T_2$ was measured by monitoring the decay of the spin-echo intensity $I(\tau)$ as a function of the interval time $\tau$ between the $\pi/2$ and $\pi$ pulses, and was evaluated by fitting $I(\tau)$ to the exponential function. We obtained satisfactory fitting by a single component of $T_1$($T_2$) for $R(t)$ ($I(\tau)$) (insets to Fig.\,3), showing that the spin fluctuations are homogeneous.

 While the $1/T_1$ and $1/T_2$ are nearly field-independent at lower fields, both quantities start to increase above $\sim 15$~T, and show a tendency to diverge above 32~T. 
As mentioned, NMR spin-echo signals were not observed above 32~T (grey area in Fig. 3 and 4). In this region, $T_2$ values become extremely short, much shorter than 3 $\mu s$, the dead time of our NMR spectrometer. This confirms the divergence of fluctuations in the vicinity of the field-induced metamagnetic transition at $H_m\approx35$~T.
Previous macroscopic studies have defined $\mu_0H^{\ast}\sim15$~T as the characteristic field above which $H_{c2}$ shows an upturn as the HFSC phase emerges on top of the LFSC phase \cite{Rosuel2022_Field-induced,sakai2022field}. 
Interestingly, the $\mu_0H^{\ast}$ has been found to remain unchanged between UTe$_2$ crystals with different qualities, even though the improvement of the crystal quality markedly increases both the onset $T_{\rm c}$ and the extrapolated $H_{\rm c2}^{\rm LFSC}(0)$  \cite{sakai2022field}.
This implies that the $H^{*}$ is not simply determined as the intersection of two SC phase boundaries, but that there are some other sources to locate the $H^{*}$ around 15~T.
Our NMR results show that the $H^{\ast}$ is defined as a characteristic field above which critical fluctuations begin to develop toward the metamagnetic transition.
It has been demonstrated \cite{Rosuel2022_Field-induced,Aoki2022Unconventional-} 
that even the rather sharp upturns observed on $H_{c2}$ could be reproduced with a smooth continuous increase of the pairing strength $\lambda(H)$, whose field dependence is related to the growth of critical fluctuations.

\begin{figure}[tb]
\begin{center}
\includegraphics[width=6.8 cm,keepaspectratio]{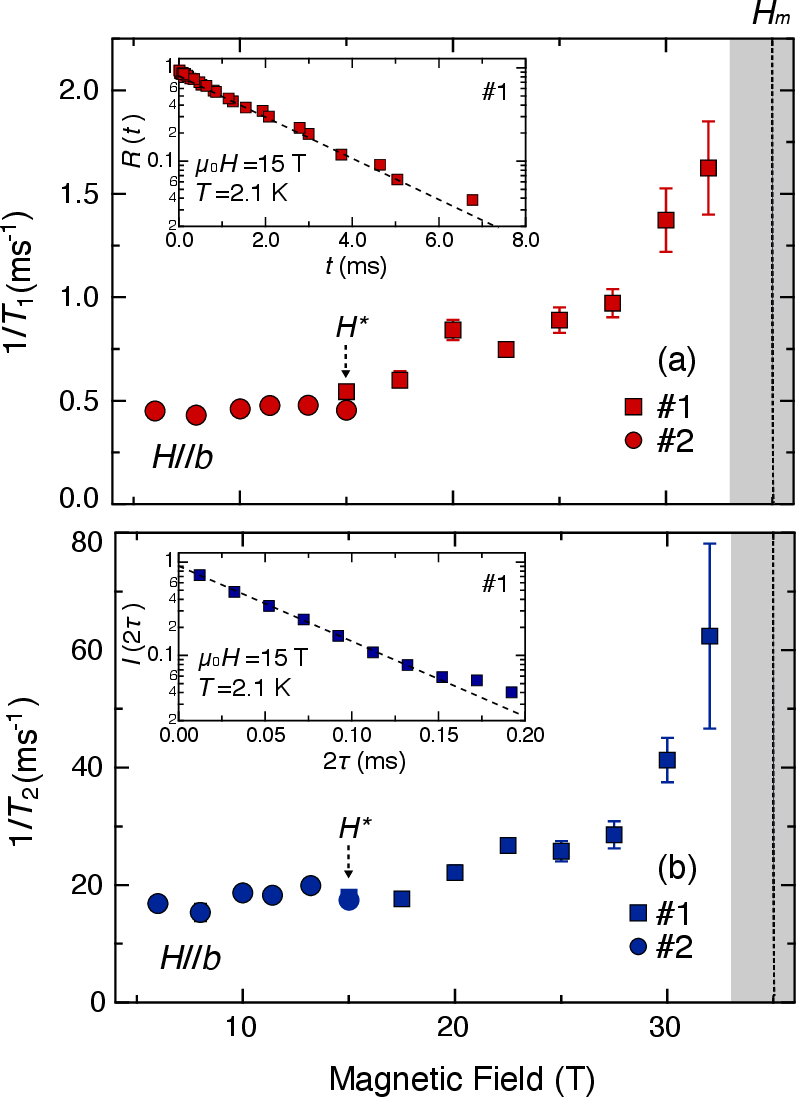} %8.5cm
\caption{The field dependence of  (a) $1/T_1$ and (b) $1/T_2$ in $H\|b$. 
In the grey area above 32~T,  we lost NMR spin-echo signals due to extremely short $T_2$. Insets display examples of the measured relaxation curves (symbols) and the exponential fits (dotted lines).
}
\end{center}
\label{T1T2}
\vspace{-8mm}
\end{figure}

In the following, we discuss the nature of fluctuations detected by $1/T_1$ and $1/T_2$. In general, $1/T_2$ is given by the sum of electronic and nuclear contributions, $1/T_2=(1/T_2)^{\rm el}+(1/T_2)^{\rm
nu}$. In the present case, we can safely assume that $(1/T_2)^{\rm
el}\gg(1/T_2)^{\rm nu}$, since $1/T_2$ values are confirmed to be independent of the $^{125}$Te isotope concentration.
$1/T_1$ and $1/T_2$ are thus both determined by electronic contributions. 
For $I = 1/2$ nuclear spins in the magnetic field applied along the $b$ direction, the electronic-spin-induced fluctuations of the local magnetic field $h(t)$ will induce the $1/T_1$ relaxation according to Moriya's formula \cite{Moriya_LocalMomentLimit}, 
\begin{equation}
1/T_{1,b}= [u_{\rm aa}(\omega_{\rm NMR}) + u_{\rm cc}(\omega_{\rm NMR})]/2,
\end{equation}
% where  
\begin{flalign}
& {\rm where}\,\,\,\,\,\,\,u_{\rm AB}(\omega) = \gamma^2 \int_{-\infty}^{+\infty}\left<h_{\rm A}(0)h_{\rm B}(t)\right>e^{-i \omega t} dt &
\end{flalign}
is the Fourier transform of the field-fluctuations correlation function. The same fluctuations also contribute to the so-called Redfield term in the $T_2$ relaxation \cite{Slichter1989,Abragam1961Principles-od-N,WalstedtRE:PRL19:1967,Horvatic2002}:
\begin{flalign}
\,\,\,\,\,\,\,1/T_{2,b} & = u_{\rm bb}(0)/2 + [u_{\rm aa}(\omega_{\rm NMR}) + u_{\rm cc}(\omega_{\rm NMR})]/4 & \nonumber \\
& = u_{\rm bb}(0)/2 + (1/T_{1,b})/2. & \label{eq3}
\end{flalign}
The first term is determined by the {\it{longitudinal}} $(\|b)$ fluctuations of $h(t)$ averaged over the time scale of $T_2$ itself. It is thus sensitive to {\it{slow}} fluctuations of electronic, magnetic moments as compared to those at the NMR frequency $\omega_{\rm NMR}$ of the order of 100~MHz. On the other hand, the second term is caused by the transverse ($\perp$$b$) fluctuations of $h(t)$ probed at $\omega_{\rm NMR}$ that define $1/T_1$.

In the absence of frequency dependence of the local field fluctuations, Eq.~(\ref{eq3}) tells us that for pure transverse, isotropic, or longitudinal fluctuations, we expect that the ratio $(1/T_{2,b})/(1/T_{1,b})$ respectively equals to 1/2, 1, or $\infty$. The experimental value $(1/T_{2,b})/(1/T_{1,b}) \approx 36$ (Fig.~4) then clearly points to dominant longitudinal fluctuations. However, if we admit possible strong frequency dependence of fluctuations, the conclusion is less evident. Nevertheless, in this case, it is clear that $1/T_2$ and $1/T_1$ would present very different $\omega_{\rm NMR} \propto H$ dependence, with $1/T_2$ being essentially field independent and $1/T_1$ {\it decreasing} with the field. This is in strong contrast to what is observed: both quantities have essentially identical, {\it increasing} field dependence (Fig. 4). We thus conclude that both quantities track the same dominant {\it longitudinal} spin fluctuations.

Precisely this will be realized if strong anisotropy of spin fluctuations overcomes the weakness of the (squared) off-diagonal terms of the hyperfine coupling tensor, $A_{ab}^2, A_{cb}^2 < A_{aa}^2, A_{cc}^2$. Here we recall that the $A$-tensor relates the electronic spin/moment to the local field felt by the nuclei \cite{Slichter1989,Abragam1961Principles-od-N}. In the Supplemental Material \cite{supple}, we show that in this case the field {\it independent} ratio $(1/T_{2,b})/(1/T_{1,b}) = A_{bb}^2/(A_{ab}^2 + A_{cb}^2) + 1/2$ is observed only when the spin fluctuations are themselves frequency independent between $\omega = 0$ and $\omega = \omega_{\rm NMR}$.

We also remark that a developement of extremely slow fluctuations, observed in $1/T_2$ but not in $1/T_1$, was detected in our previous study \cite{Tokunaga2022Slow-Electronic} using a CVT crystal of UTe$_2$ at low fields, in $H\|a$ but not in $H\|b$. A low temperature upturn in the spin susceptibility ($H\|a$) is attributed to disorder or defects that are intrinsically present in CVT crystals, but do not exist in the presently studied MSF crystal \cite{Sakai2022Single-crystal-}.

In Fig.\,4,  we plot the $H$-dependent quadratic coefficient $A(H)$ together with our $1/T_1$ and $1/T_2$ data.
Here the $A(H)$ is extracted from a Fermi-liquid fit to the electrical resistivity at low temperatures (with current $I\|a$) by $\rho(T)=\rho_0+AT^2$ in $H\|b$ \cite{Knafo2019Magnetic-Field-,Knafo2021Comparison-of-t,Thebault_UTe2_2022}. The significant increase in $A(H)$ near $H_{\rm m}$ has been regarded as due to the enhancement of the effective mass $m^*$ \cite{Knafo2019Magnetic-Field-,Knafo2021Comparison-of-t,Thebault_UTe2_2022}.
We found a reasonable scaling between the $A(H)$ and the NMR relaxation rates up to 32~T.
A similar strong field-dependence, accompanying a sharp maximum around $H_{\rm m}$, has also been observed in the Sommerfeld coefficient $\gamma (H)$, which was derived in literature from the specific-heat data at low $T$ \cite{Imajo2019Thermodynamic-I,Rosuel2022_Field-induced} or the $T$-dependence of magnetization using a thermodynamic
Maxwell relation \cite{Miyake2019Metamagnetic-Tr,Miyake_UTe2_2021}.
Our NMR experiments demonstrate the importance of longitudinal spin fluctuations in the entire field range and that their development underlies these unconventional transport and thermal behaviors in $H\|b$.
Remarkably, Miyake has predicted based on an extended Landau theory of the first-order metamagnatic transition that enhanced longitudinal spin fluctuations would give rise to the enhancement of the $A$ coefficient and the $\gamma$ around the metamagnetic field \cite{doi:10.7566/JPSJ.90.024701}.

\begin{figure}[tb]
\begin{center}
\includegraphics[width=7.4 cm,keepaspectratio]{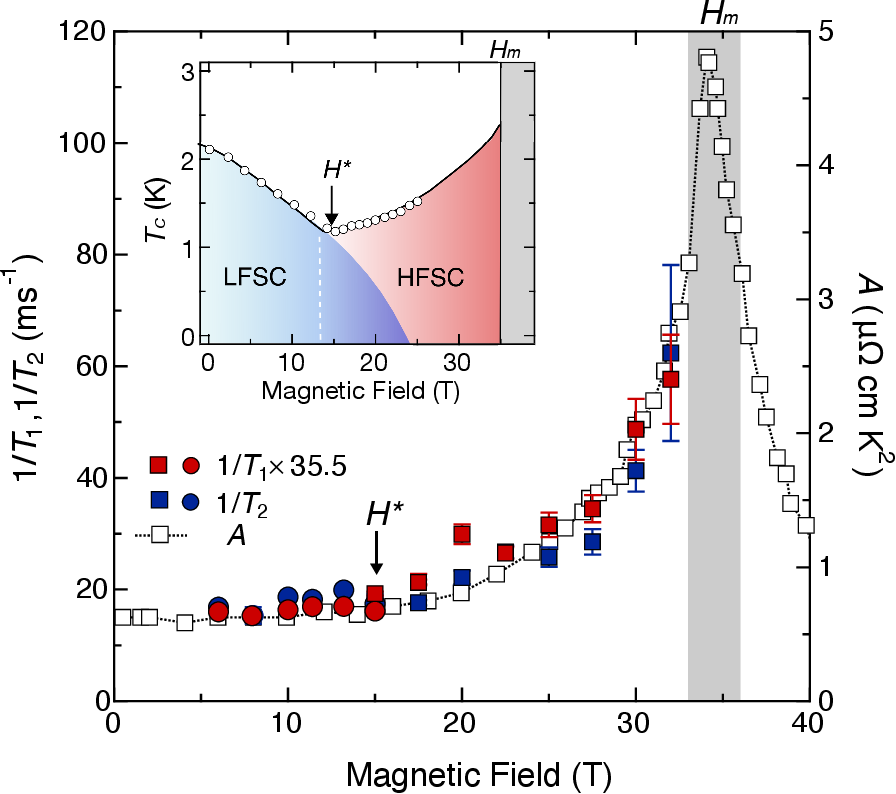} %8.5cm
\caption{The field dependent quadratic coefficient $A(H)$ (with current $I\|a$)  reported in Ref. \cite{Knafo2019Magnetic-Field-} is compared with $1/T_2(H)$ and with  $1/T_1(H)$ scaled by a factor of 35.5. 
The inset shows a schematic $H-T$ phase diagram for UTe$_2$ in the case of $H\|b$ \cite{Rosuel2022_Field-induced,Kinjo2022_changeof,sakai2022field}. The $T_{\rm c}(H)$ data (white circle) is from Ref. \cite{sakai2022field}.
}
\end{center}
\label{AvsH}
\vspace{-8mm}
\end{figure}

It should be noted that the diverging $1/T_2$ behavior observed here in UTe$_2$ is very similar to that in URhGe.
The latter compound is a ferromagnet with a strong Ising-type spin anisotropy and exhibits a metamagnetic-like transition from a FM state to a high-field polarized state when a field of $H_{\rm m}=12$~T is applied along a hard-magnetization axis ($\|b$). The diverging $1/T_2$ has been observed near the critical fields, which is also attributed to the divergence of longitudinal spin fluctuations \cite{Tokunaga2015Reentrant-Super,TokunagaY:PRB93:2016,TokunagaY:JPSCP30:2020, KotegawaH:JPSJ84:2015}. 
For URhGe, the divergence has been associated with the presence of a tricritical point (TCP).
The TCP locates at 12~T and around 5~K \cite{Levy1343,Levy:2007aa,Nakamura2017_Wing,Gourgout_PRL_2016}, below which a second-order transition changes to a first-order transition. With regard to magnetic excitations, 
a characteristic feature of the TCP is that it can trigger a diverging susceptibility not only for the order parameter ($M_{\rm c}$) but also for the physical quantity that is conjugate to the tuning parameter $H_{\rm b}$ driving the phase transition, that is $M_{\rm b}$ in the case of URhGe \cite{Huxley2007Odd,Misawa2006,Misawa2008,Tokunaga2015Reentrant-Super,TokunagaY:PRB93:2016}. 
Thus, we can expect a divergence of the longitudinal component of magnetic fluctuations, resulting in the diverging $1/T_2$ in $H\|b$.
In URhGe, the reentrant SC has been found to occur in almost the same region as that where the longitudinal fluctuations are developing \cite{Tokunaga2015Reentrant-Super,TokunagaY:PRB93:2016}.
On the other hand, there is no TCP in UCoGe, and hence, only a broad peak appears in $1/T_2$ around a critical field of $12-13$~T \cite{Ishida2021Pairing-interac}.

For the metamagnetic transition in UTe$_2$,  there is no TCP, but a critical
point (CP) exists at $T=5-7$~K with $H_{\rm m}=35$~T \cite{Miyake2019Metamagnetic-Tr,Knafo2019Magnetic-Field-,Knafo2021Comparison-of-t}.
Below the CP, a metamagnetic crossover at higher temperatures changes to a first-order transition.
In this context, it is also important to notice that  the diverging $1/T_2$ has also been observed in the vicinity of the CP (at 10~K, 1~T) in an itinerant paramagnet UCoAl \cite{Nohara2011,Karube2015}. 
This compound is located on the paramagnetic side in the vicinity of FM order and a metamagnetic transition to a FM phase occurs under a small $H$ ($\sim1$ T). 
In the case of UCoAl, however, the critical fluctuations are suppressed rapidly in the first-order region below 10~K, making clear contrast to the case of UTe$_2$. This might be related to the fact that  the CP of UCoAl occurs under $H$ applied along the easy-magnetization axis \cite{Aoki2011Feeromagnetic,Aoki2022Unconventional-}. It is also remarkable that both UTe$_2$ and UCoAl exhibit a broad maximum in $\chi(T)$ around $20-30$~K \cite{Aoki2011Feeromagnetic, Ran2019Nearly-ferromag}.

Theoretically, spin fluctuations near a FM quantum critical point have been supposed to create a binding force between quasiparticles with equal (triplet) spin pairs \cite{fay1980coexistence,valls1984superconductivity}, analogous to the mechanism of a superfluid pairing in $^{3}$He \cite{Levin_PRB_1978}. After the discovery of the uranium-based FM superconductors, theoretical models of the spin-triplet SC under magnetic fields were developed by Mineev \cite{Mineev2011Magnetic-field-,Mineev_PRB_2015,PhysRevB.103.144508}, Tada and Fujimoto \cite{Tada_2013,Tada2016}, and Hattori and Tsunetsugu \cite{Hattori_Tsunetsugu_PRB_2013}, independently. Those theories have pointed out the importance of field-dependent spin susceptibility and, thus, spin fluctuations. More recently, using the density matrix renormalization group, Suzuki and Hattori have analyzed SC correlations in the one-dimensional Kondo lattice models with Ising anisotropy under transverse magnetic fields \cite{SuzukiHattori2019,Hattori_Suzuki_2020}.
They found that competitions between the transverse magnetic field and the Kondo singlet formation lead to both enhanced SC correlations and metamagnetic  behaviors, where metamagnetic fluctuations play a crucial role \cite{Hattori_Suzuki_2020}.
The present NMR experimentally captures the fluctuations that well corroborate these theoretical models, although a more quantitative comparison between experiments and theories remains for future works.
We will also extend the NMR experiments to higher magnetic fields above $H_{\rm m}$, where the rapid suppression of the diverging fluctuations is expected, accompanied by the sudden decrease of $T_{\rm c}(H)$.

%%%%%%%%%%%%%%%%%%%%%%%%%%%%%%%%%%%%%%%%%%%%%%%%%%%%%%%%%%%%%%%%%%%%%%%%%%
%%%%      Acknowledgments
%%%%%%%%%%%%%%%%%%%%%%%%%%%%%%%%%%%%%%%%%%%%%%%%%%%%%%%%%%%%%%%%%%%%%%%%%%
\vspace{2em}
\begin{acknowledgments}
We are grateful for the stimulating discussions with J.P. Brison, W. Knafo, D. Braithwaite, A. Pourret, J. Flouquet, M.-H. Julien, H. Mayaffre, I. Sheikin, K. Machida, K. Miyake and Y. Yanase.
This work (a part of high magnetic field experiments) was performed at LNCMI under the EMFL program (Proposal Numbers GMS01-122, SG-MA0222.). A part of this work was supported by the French National Agency for Research ANR within the
project FRESCO No. ANR-20-CE30-0020 and FETTOM No. ANR-19-CE30-0037 and by JSPS KAKENHI Grant Nos. JP16KK0106,  JP20H00130, JP20KK0061, JP20K20905, JP22H04933, JP23H01132, JP23H04871, JP23K03332, and JP23H01124 and by the JAEA REIMEI Research Program.
\end{acknowledgments}

\newpage

\title{Supplementary Information for ``Longitudinal spin fluctuations driving field-reinforced superconductivity in UTe$_2$"}


% \author
% \email[]{Your e-mail address}
% \homepage[]{Your web page}
% \thanks{}
% \altaffiliation{}
% \affiliation{}

%%%%%%%%%%%%%%%%%%%%%%%%%%%%%%%%%%%%%%%%%%%%%%%%%%%%%%%%%%%
\author{Y. Tokunaga}
%\email[]{tokunaga.yo@jaea.go.jp}
\affiliation{Advanced Science Research Center, Japan Atomic Energy Agency, Tokai, Ibaraki 319-1195, Japan}

\author{H. Sakai}
\affiliation{Advanced Science Research Center, Japan Atomic Energy Agency, Tokai, Ibaraki 319-1195, Japan}

\author{S. Kambe}
\affiliation{Advanced Science Research Center, Japan Atomic Energy Agency, Tokai, Ibaraki 319-1195, Japan}

\author{P. Opletal}
\affiliation{Advanced Science Research Center, Japan Atomic Energy Agency, Tokai, Ibaraki 319-1195, Japan}

\author{Y. Tokiwa}
\affiliation{Advanced Science Research Center, Japan Atomic Energy Agency, Tokai, Ibaraki 319-1195, Japan}

\author{Y. Haga}
\affiliation{Advanced Science Research Center, Japan Atomic Energy Agency, Tokai, Ibaraki 319-1195, Japan}
%%%%%%%%%%%%%%%%%%%%%%%%%%%%%%%%%%%%%%%%%%%%%%%%%%%%%%%%%%%

%%%%%%%%%%%%%%%%%%%%%%%%%%%%%%%%%%%%%%%%%%%%%%%%
\author{S. Kitagawa}
\affiliation{Department of Physics, Graduate School of Science, 
Kyoto University, Kyoto 606-8502, Japan}

\author{K. Ishida}
\affiliation{Department of Physics, Graduate School of Science, 
Kyoto University, Kyoto 606-8502, Japan}
%%%%%%%%%%%%%%%%%%%%%%%%%%%%%%%%%%%%%%%%%%%%%%%%

%%%%%%%%%%%%%%%%%%%%%%%%%%%%%%%%%%%%%%%%%%%%%%%%
\author{D. Aoki}
\affiliation{IMR, Tohoku University, Ibaraki 311-1313, Japan}
\affiliation{%
Univ. Grenoble Alpes, CEA, Grenoble-INP, IRIG, Pheliqs, 38000 Grenoble France
}%
%%%%%%%%%%%%%%%%%%%%%%%%%%%%%%%%%%%%%%%%%%%%%%%%

%%%%%%%%%%%%%%%%%%%%%%%%%%%%%%%%%%%%%%%%%%%%%%%%
\author{G.\,Knebel}
\affiliation{%
Univ. Grenoble Alpes, CEA, Grenoble-INP, IRIG, Pheliqs, 38000 Grenoble France
}%
\author{G. \,Lapertot}
\affiliation{%
Univ. Grenoble Alpes, CEA, Grenoble-INP, IRIG, Pheliqs, 38000 Grenoble France
}
%%%%%%%%%%%%%%%%%%%%%%%%%%%%%%%%%%%%%%%%%%%%%%%%
\author{S.\,Kr\"{a}mer}
\affiliation{%
LNCMI-CNRS (UPR3228), EMFL, UGA-UPS-INSA, 38042 Grenoble, France
}%
\author{M.\,Horvati{\'c}}
\affiliation{%
LNCMI-CNRS (UPR3228), EMFL, , 38042 Grenoble, France
}%


\date{\today}

%\begin{abstract}
%\end{abstract}

%\keywords{}

\maketitle


%
\section{\NoCaseChange{The $1/T_1$ and the Redfield $1/T_2$ NMR rates for $I=1/2$ and magnetic fluctuations}}
For nuclear $I= 1/2$ spins subject to magnetic fluctuations, we discuss the nuclear spin-lattice relaxation rate $1/T_1$ and the corresponding Redfield contribution to the nuclear spin-spin relaxation rate $1/T_2$. 
When the fluctuations of the local magnetic field at the nuclear site are described by
\begin{equation}
u_{\rm AB}(\omega)=\gamma^2 \int_{-\infty}^{+\infty}\left<h_{\rm A}(0)h_{\rm B}(t)\right>e^{-i \omega t} dt ,
\end{equation}
where A and B denote one of the coordinate axes $x, y, z$, in the external magnetic field applied along the $z$ direction, the $1/T_1$ rate is given by the Moriya's formula \cite{Moriya_LocalMomentLimitSM}
\begin{equation}
1/T_{1,z} = [u_{xx}(\omega_{\rm NMR})+u_{yy}(\omega_{\rm NMR})]/2,
\end{equation}
and the corresponding Redfield $1/T_2$ rate is \cite{Slichter1989SM,Horvatic2002SM}
\begin{equation}
\begin{split}
1/T_{2,z}&= u_{zz}(0)/2 + [u_{xx}(\omega_{\rm NMR})+u_{yy}(\omega_{\rm NMR})]/4\\
&= u_{zz}(0)/{2}+(1/T_{1,z})/2.
\end{split}
\end{equation}
In terms of the corresponding electron spin fluctuations
\begin{equation}
G_{\rm AB}(\omega)=\gamma^2 \int_{-\infty}^{+\infty}\left<s_{\rm A}(0)s_{\rm B}(t)\right>e^{-i \omega t} dt,
\end{equation}
which are linearly related to the local field by the hyperfine coupling tensor $\overrightarrow{h}(t) = -\overleftrightarrow{A}\overrightarrow{s}(t)$, the above-given expressions become more complicated because of the presence of the off-diagonal terms $A_{xz}, A_{yz}, A_{xy}$. For example, 
\begin{equation}
\begin{split}
1/T_{1,z}&=\gamma^2(A_{xx}^2+A_{xy}^2) G_{xx}(\omega_{\rm NMR})/2\\
&+\gamma^2(A_{yy}^2+A_{xy}^2) G_{yy}(\omega_{\rm NMR})/2\\
&+\gamma^2(A_{xz}^2+A_{yz}^2) G_{zz}(\omega_{\rm NMR})/2,
\end{split}
\end{equation}
where $x, y, z$ are considered to be the principal axes of the $G_{AB}(\omega)$ tensor, to ensure that it is diagonal.

In the following we apply these formulas to UTe$_2$ in the magnetic field applied along the $b$ axis ($z = b$). Since there $1/T_{2,b} \gg 1/T_{1,b}$, from Eqs. (1-3) we directly conclude that $u_{bb}(0) \gg u_{aa}(\omega_{\rm NMR})$, $u_{cc}(\omega_{\rm NMR})$. The $\overleftrightarrow{A}$ tensor has only a weak anisotropy, $A_{bb}, A_{cc}, A_{aa}$~= 5.2, 3.9, 4.7~T/$\mu_{B}$ \cite{Fujibayashi2023Low-TemperatureSM}, which implies that its off-diagonal terms---in general of the same size as the anisotropy of the diagonal terms---are negligible. This in turn implies that the above condition will also be valid for the spin fluctuations: $G_{bb}(0)\gg G_{aa}(\omega_{\rm NMR}), G_{cc}(\omega_{\rm NMR})$.

If the strength of the longitudinal fluctuations, \mbox{$G_{bb} \gg G_{aa}, G_{cc}$}, overcomes the weakness of the off-diagonal coupling, $A_{ab}^2, A_{cb}^2 < A_{aa}^2, A_{cc}^2$, then in Eq. (5) the last term is dominant
\begin{equation}
1/T_{1,b} \simeq \gamma^2(A_{ab}^2+A_{cb}^2) G_{bb}(\omega_{\rm NMR})/2,
\end{equation}
and both $1/T_1$ and $1/T_2$ are dominated by the longitudinal fluctuations:
\begin{equation}
1/T_{2,b}=\gamma^2A_{bb}^2 G_{bb}(0)/2+(1/T_{1,b})/2.
\end{equation}
As their ratio
\begin{equation}
\frac{1/T_{2,b}}{1/T_{1,b}}=\frac{A_{bb}^2}{A_{ab}^2+A_{cb}^2} \frac{G_{bb}(0)}{G_{bb}(\omega_{\rm NMR})}+\frac{1}{2}
\end{equation}
is experimentally field independent (Fig.~4), the longitudinal spin fluctuations must be frequency independent in the corresponding frequency range, 
$G_{bb}(0) = G_{bb}(\omega_{\rm NMR})$, so that the ratio 
\begin{equation}
\frac{1/T_{2,b}}{1/T_{1,b}}=\frac{A_{bb}^2}{A_{ab}^2+A_{cb}^2}+\frac{1}{2}
\end{equation}
is defined only by the ratio of the $b$-axis diagonal and off-diagonal couplings. The experimental value of the ratio $(1/T_{2,b})/(1/T_{1,b}) \approx 35.5$ then implies that the RMS average of the two off-diagonal coupling terms $A_{ab}, A_{cb}$ equals 0.8~T/$\mu_{B}$, which is comparable to the anisotropy of the diagonal couplings.

\end{document}